\documentclass[a4paper,11pt,fleqn]{article}

\usepackage{hyperref}
\usepackage[ansinew]{inputenc}
\usepackage[mathscr]{eucal}
\usepackage{amsmath,amssymb,amsthm}
\usepackage{graphicx}
\usepackage{cite}

\allowdisplaybreaks

\hypersetup{colorlinks, linkcolor=blue, citecolor=blue, urlcolor=blue}
\newcommand{\p}{\partial}
\newcommand{\const}{\mathop{\rm const}\nolimits}

\newcommand{\todo}[1][\null]{\ensuremath{\clubsuit}}

\newcommand{\noprint}[1]{}

\newtheorem{theorem}{Theorem}

{\theoremstyle{definition}

\newtheorem{remark}{Remark}
\newtheorem*{remark*}{Remark}
}

\newcommand{\checked}[1][\null]{\ensuremath{\boldsymbol{\surd}}}

\newcommand{\DD}{\mathrm{D}}

\newcommand{\vv}{\mathbf{v}}

\newcommand{\ve}{\varepsilon}

\newcommand{\DDD}{\mathcal{D}}
\newcommand{\GGG}{\mathcal{G}}
\newcommand{\LLL}{\mathcal{L}}

\begin{document}

\par\noindent {\LARGE\bf
Symmetry analysis of a system of \\
modified shallow-water equations
\par}

{\vspace{4mm}\par\noindent {\large Simon Szatmari$^\dag$ and Alexander Bihlo$\hspace{0.1mm}^\ddag$
} \par\vspace{2mm}\par}

{\vspace{2mm}\par\noindent {\it
$^{\dag}$Department of Mathematics and Statistics, McGill University, 805 Sherbrooke W.,\\
$\phantom{^\dag}$Montr\'{e}al (QC) H3A 2K6, Canada\\
}}
{\noindent \vspace{2mm}{\it
$\phantom{^\dag}$\textup{E-mail}: simon.szatmari@mail.mcgill.ca
}\par}

{\vspace{2mm}\par\noindent {\it
$^{\ddag}$Centre de recherches math\'{e}matiques, Universit\'{e} de Montr\'{e}al, C.P.\ 6128, succ.\
Centre-ville,\\$\phantom{^\dag}$Montr\'{e}al (QC) H3C 3J7, Canada\\
}}
{\noindent \vspace{2mm}{\it
$\phantom{^\dag}$\textup{E-mail}: bihlo@crm.umontreal.ca
}\par}

\vspace{2mm}\par\noindent\hspace*{5mm}\parbox{150mm}{\small
We revise the symmetry analysis of a modified system of one-dimensional shallow-water equations (MSWE) recently
considered by Raja Sekhar and Sharma [\textit{Commun.\ Nonlinear Sci.\ Numer.\ Simulat.} \textbf{20} (2012)
630--636]. Only a finite dimensional subalgebra of the maximal Lie invariance algebra of the MSWE, which in fact is
infinite dimensional, was found in the aforementioned paper. The MSWE can be linearized using a hodograph
transformation. An optimal list of inequivalent one-dimensional subalgebras of the maximal Lie invariance algebra is
constructed and used for Lie reductions. Non-Lie solutions are found from solutions of the linearized MSWE.
}\par\vspace{2mm}

\section{Introduction}

The computation of exact solutions of nonlinear systems of partial differential equations remains an important task,
despite the increased interest in the numerical simulations of such differential equations. Exact solutions are
crucial, for instance, in the testing of numerical methods as well as in the analytical study of the associated partial
differential equations. Finding exact solutions remains rather challenging, unless the equations of interest belong to a
category for which well-known solving methods exist. However, in general, the most one can do is to construct particular
solutions. Particular solutions, albeit often simple, can be of great help to gain understanding and insight into the
principal dynamics of such systems of equations.

One way to obtain exact solutions of differential equations is through the study of their Lie symmetries. It is worth to
know that there is also a number of non-Lie methods for finding exact solutions, see
e.g.~\cite{blum10Ay,gala06Ay,mele05Ay}. Lie symmetries and the related computational methods for finding group-invariant
solutions of systems of differential equations constitute a well investigated subject. The current paper concerns itself
with a hydrodynamical problem. Relevant results on symmetries and exact solutions for hydrodynamical systems can be
found e.g.\ in the textbooks~\cite{ibra94Ay,andr98Ay,blum89Ay,ibra95Ay,mele05Ay,olve86Ay,ovsi82Ay} and in the
papers~\cite{andr88Ay,ches09Ay,fush94Ay,hema02Ay,huar10Ay,levi89By,mele04Ay,popo99Ay,popo00Ay,popo95Ay}.

The recent paper~\cite{sekh12Ay} intended to find Lie symmetries and exact solutions of the following system of modified
one-dimensional shallow-water equations:
\begin{equation}\label{eq:ModifiedShallowWater}
 \Delta_1:=u_t+uu_x+g\left(1+\frac{H}{h}\right)h_x=0,\quad \Delta_2:=h_t+uh_x+hu_x=0,
\end{equation}
where $u$ is the fluid velocity in $x$-direction, $h$ is the height of the water column, and $g$ and $H$ are constants
related to the gravity acceleration and momentum transport, respectively. The constant $g$ can be scaled to $g=1$ by an
equivalence transformation. In the case of $H=0$, this system reduces to the usual form of the one-dimensional
shallow-water equations.

We found a number of incorrect and incomplete results in~\cite{sekh12Ay}, which is why we reconsider the symmetry
analysis of the system~\eqref{eq:ModifiedShallowWater}. One main concern is that the maximal Lie invariance algebra
of~\eqref{eq:ModifiedShallowWater} is supposed to be infinite dimensional. At the same time, in~\cite{sekh12Ay} it was
found to be finite dimensional. This sits ill with the fact that system~\eqref{eq:ModifiedShallowWater} can be
linearized via a hodograph transformation and thus should display an infinite dimensional maximal Lie invariance
algebra. Another problem with the Lie reductions carried out in~\eqref{eq:ModifiedShallowWater} is that no optimal list
of inequivalent subalgebras was determined which should be the base for an efficient computation of group-invariant
solutions. As a consequence, the exact solutions found in~\cite{sekh12Ay} are in an overly complicated form. Moreover,
we explicitly show that one of the reductions in~\cite{sekh12Ay} is incorrect.

The structure of this paper is as follows. In Section~\ref{sec:LieSymmetriesModifiedShallowWater} we compute the Lie
symmetries admitted by system~\eqref{eq:ModifiedShallowWater}. We find the infinite dimensional maximal Lie invariance
algebra of system~\eqref{eq:ModifiedShallowWater} and compare it to the algebra found in~\cite{sekh12Ay} which is only
finite dimensional. We indicate how the modified shallow-water equations are linearized through a hodograph
transformation. In Section~\ref{sec:ClassificationSubalgebrasModifiedShallowWater} we classify inequivalent
one-dimensional subalgebras of the infinite dimensional maximal Lie invariance algebra of the modified shallow-water
equations. One-dimensional subalgebras of the symmetry algebra found in~\cite{sekh12Ay} are also classified. In
Section~\ref{sec:ExactSolutionsModifiedShallowWater} we construct the group-invariant solutions based on the optimal
list found in Section~\ref{sec:ClassificationSubalgebrasModifiedShallowWater}. We also determine non-Lie solutions of
the modified shallow-water equations through solutions of the linearized modified shallow-water equations. The results
obtained are summed up in the final Section~\ref{sec:ConclusionModifiedShallowWater}.

\section{Lie symmetries of the modified shallow-water equations}\label{sec:LieSymmetriesModifiedShallowWater}

In this section, we compute the maximal Lie invariance algebra $\mathfrak g$ of the
system~\eqref{eq:ModifiedShallowWater}. This is done by determining the coefficients of the infinitesimal
generator
\[
 Q=\tau(t,x,u,h)\p_t+\xi(t,x,u,h)\p_x+\eta(t,x,u,h)\p_u+\phi(t,x,u,h)\p_h,
\]
of a general one-parameter Lie symmetry group. These coefficients are found using the infinitesimal invariance
criterion,
which in the present case reads
\begin{equation}\label{eq:InfinitesimalInvarianceCriterionModifiedShallowWater}
 Q^{(1)}\Delta_1=0,\quad Q^{(1)}\Delta_2=0,
\end{equation}
where this equality has to hold on the manifold defined by $\Delta_1=0$ and
$\Delta_2=0$. In system~\eqref{eq:InfinitesimalInvarianceCriterionModifiedShallowWater}, $Q^{(1)}$ denotes the first
prolongation of the operator $Q$, which is of the form
\[
 Q^{(1)}=Q+\eta^t\p_{u_t}+\eta^x\p_{u_x}+\phi^t\p_{h_t}+\phi^x\p_{h_x}.
\]
Here the coefficients are given by
\begin{align}\label{eq:CoefficientsNeededForProlongation}
\begin{split}
 &\eta^t=\DD_t\eta-u_t\DD_t\tau-u_x\DD_t\xi,\quad
 \eta^x=\DD_x\eta-u_t\DD_x\tau -u_x\DD_x\xi,\\
 &\phi^t=\DD_t\phi-h_t\DD_t\tau-h_x\DD_t\xi,\quad
 \phi^x=\DD_x\phi-h_t\DD_x\tau-h_x\DD_x\xi,
\end{split}
\end{align}
which follow from the \emph{standard prolongation formula}, see
e.g.~\cite{andr98Ay,blum89Ay,blum10Ay,olve86Ay,ovsi82Ay}. The total derivative operators $\DD_t$ and $\DD_x$ arising in
the expressions of the coefficients $\eta^t$,
$\eta^x$, $\phi^t$ and $\phi^x$ of the prolonged operator $Q^{(1)}$ are given by
\begin{align*}
 &\DD_t=\p_t+u_t\p_u+h_t\p_h+u_{tt}\p_{u_t}+h_{tt}\p_{h_t}+u_{tx}\p_{u_x}+h_{tx}\p_{h_x}+\cdots, \\
 &\DD_x=\p_x+u_x\p_u+h_x\p_h+u_{tx}\p_{u_t}+h_{tx}\p_{h_t}+u_{xx}\p_{u_x}+h_{xx}\p_{h_x}+\cdots.
\end{align*}
Explicitly, the infinitesimal invariance criterion~\eqref{eq:InfinitesimalInvarianceCriterionModifiedShallowWater} reads
\begin{align*}
 &\eta^t+u\eta^x+\eta u_x+\left(1+\frac{H}{h}\right)\phi^x-\frac{H}{h^2}\phi h_x=0,\\
 &\phi^t+u\phi^x+\eta h_x+ h\eta^x+\phi u_x=0.
\end{align*}
Plugging the coefficients~\eqref{eq:CoefficientsNeededForProlongation} into the above system, substituting
$u_t=-(uu_x+(1+H/h)h_x)$ and $h_t=-(uh_x+hu_x)$ wherever they arise and splitting the resulting equations with respect
to powers of the derivatives of $u$ and $h$, we derive the following system of determining equations for coefficients of
the vector field $Q$:
\begin{align}\label{eq:DeterminingEquationsModifiedShallowWater}
\begin{split}
 &\xi_u-u\tau_u+h\tau_h=0,\\
 &\xi_h-u\tau_h+\left(1+\frac{H}{h}\right)\tau_u=0,\\
 &\frac{H}{h^2}\phi-\left(1+\frac{H}{h}\right)(\tau_t-\xi_x-\eta_u+\phi_h+2u\tau_x)=0,\\
 &\phi+h(\tau_t-\xi_x+\eta_u-\phi_h+2u\tau_x)=0,\\
 &\eta-h\eta_h+u(\tau_t-\xi_x)-\xi_t+u^2\tau_x+\left(1+\frac{H}{h}\right)(\phi_u+h\tau_x)=0,\\
 &\eta+h\eta_h+u(\tau_t-\xi_x)-\xi_t+u^2\tau_x-\left(1+\frac{H}{h}\right)(\phi_u-h\tau_x)=0,\\
 &\eta_t+u\eta_x+\left(1+\frac{H}{h}\right)\phi_x=0,\\
 &\phi_t+u\phi_x+h\eta_x=0.
\end{split}
\end{align}
This system has two inequivalent cases for solutions, depending on whether $H\ne0$ or $H=0$. 

If $H=0$, system~\eqref{eq:ModifiedShallowWater} reduces to the usual one-dimensional shallow-water equations for which
the general solution of the determining equations~\eqref{eq:DeterminingEquationsModifiedShallowWater} is
\begin{align*}
 &\tau=c_1t+c_4(2x-6tu)+f(u,h),\quad \xi=c_1x+c_2t+c_3x+c_4t(6h-3u^2)+g(u,h),\\ &\eta=c_2+c_3u+c_4(u^2+4h),\quad
\phi=2c_3h+4c_4uh,
\end{align*}
where $c_1,\dots,c_4$ are arbitrary reals constants and the functions $f$ and $g$ run through the set of solutions of
the system
\[
 g_u-uf_u+hf_h=0,\quad g_h-uf_h+f_u=0.
\]
This is the linearized system of one-dimensional shallow-water equations, obtained from the shallow-water equations
through the hodograph transformation in which $f=t$ and $g=x$ serve as the new unknown functions and $u$ and $h$ become
the new independent variables. The maximal Lie invariance algebra of the one-dimensional shallow-water equations is thus
spanned by the vector fields
\begin{align*}
&\DDD_1 = t\p_t+x\p_x,\quad \GGG = t\p_x+\p_u,\quad \DDD_2 = 2h\p_h + u\p_u - t\p_t,\\
&\mathcal{C} = 4hu\p_h + (4hg+u^2)\p_u + (-6ut+2x)\p_t + (6hgt-3u^2 t)\p_x, \\
&\LLL (f,g)=f(u,h)\p_t+g(u,h)\p_x,
\end{align*}
see also~\cite{bihl12By,hydo00Ay}.

We will only concentrate on the case $H\neq0$ subsequently, which corresponds to the modified shallow-water equations.
If $H \neq 0$, then the system of
determining equations~\eqref{eq:DeterminingEquationsModifiedShallowWater} yields the general solution
\[
 \tau=c_1t+f(u,h),\quad \xi=c_1x+c_2t+g(u,h),\quad \eta=c_2,\quad \phi=0,
\]
where $c_1$ and $c_2$ are arbitrary real constants and the functions $f$ and $g$ run through the set of solutions of the
system
\begin{equation}\label{eq:LinearizationModifiedShallowWater}
 g_u-uf_u+hf_h=0,\quad g_h-uf_h+\left(1+\frac{H}{h}\right)f_u=0.
\end{equation}
The maximal Lie invariance algebra $\mathfrak g$
of system~\eqref{eq:ModifiedShallowWater} with $H\ne0$ is also infinite dimensional and spanned by the vector fields
\begin{equation}\label{eq:MaximalLieInvarianceAlgebraModifiedShallowWater}
 \DDD=t\p_t+x\p_x,\quad \GGG=t\p_x+\p_u,\quad \LLL(f,g)=f(u,h)\p_t+g(u,h)\p_x.
\end{equation}
The first operator is associated with scaling symmetries in $t$ and $x$ and the second operator gives rise to Galilean
boosts. Similarly to the usual one-dimensional shallow-water equations, the operator $\LLL(f,g)$ indicates that
system~\eqref{eq:ModifiedShallowWater} can be linearized to system~\eqref{eq:LinearizationModifiedShallowWater} by the
hodograph transformation in which $f=t$ and $g=x$ are the new unknown functions of the new independent variables $u$ and
$h$. 

We have verified our computations, leading to the maximal Lie invariance algebra $\mathfrak g$, using the
package \texttt{DESOLV} for finding Lie symmetries~\cite{vu12Ay}.

Besides Lie symmetries, also discrete point symmetries of systems of differential equations can be useful, especially in
the classification of optimal lists of inequivalent subalgebras or the construction of structure-preserving
low-dimensional dynamical systems, see e.g.~\cite{bihl09Ay,bihl11Ay}. Discrete point symmetries can be computed
systematically using the direct method or, often more convenient, a version of the algebraic method proposed
in~\cite{hydo00By} and refined in~\cite{bihl11Cy,card12Ay} for systems admitting infinite dimensional maximal Lie
invariance algebras. In the present case, it is found that system~\eqref{eq:ModifiedShallowWater} admits two independent
discrete point symmetries (up to composition with continuous symmetries and each other), which are given by the pairwise
alternation of signs in $t$, $x$ and $x$, $u$, respectively.

\begin{remark}
 In~\cite{sekh12Ay} the authors stated that the maximal Lie invariance algebra of system~\eqref{eq:ModifiedShallowWater}
is spanned by the four vector fields
 \[
  \p_t,\quad \p_x,\quad t\p_x+\p_u,\quad t\p_t+x\p_x.
 \]
 Comparing this algebra, which we denote by $\mathfrak g^1$, with the algebra spanned the the vector
fields~\eqref{eq:MaximalLieInvarianceAlgebraModifiedShallowWater}, it is obvious that $\mathfrak g^1$ is only a
subalgebra of $\mathfrak g$ since the first two vector fields are special instances of the parameterized vector field
$\LLL(f,g)$, for which $(f,g)=(1,0)$ and $(f,g)=(0,1)$, respectively. It is thus also missed in~\cite{sekh12Ay} that
system~\eqref{eq:ModifiedShallowWater} can be linearized using a hodograph transformation.
\end{remark}

In the following, we denote by $G^1$ the subgroup of the maximal Lie invariance pseudogroup $G$ of the modified
shallow-water equations~\eqref{eq:ModifiedShallowWater} that is associated with the subalgebra $\mathfrak g^1$.

\section{Classification of subalgebras}\label{sec:ClassificationSubalgebrasModifiedShallowWater}

The computation of exact solutions of system~\eqref{eq:ModifiedShallowWater} was of central importance
in~\cite{sekh12Ay}. This was done with the method of group-invariant (or Lie) reduction. These reductions
were constructed without a classification of optimal lists of inequivalent subalgebras. Unfortunately such
classifications are commonly omitted in the literature without justification. This topic is extensively discussed in
e.g.~\cite{olve86Ay,ovsi82Ay}, where it is pointed out that only group-invariant solutions stemming from inequivalent
subalgebras of the maximal Lie invariance algebra of a system of differential equations are guaranteed to be
inequivalent. This means that they cannot be related to each other by means of a symmetry transformation.

Hence, the first step in the construction of exact solutions of a system of differential equations using Lie reduction
should be the classification of subalgebras of appropriate dimensions. In the present case,
system~\eqref{eq:ModifiedShallowWater} is a system in $(1+1)$ dimensions, so it is sufficient to carry out the Lie
reductions with respect to one-dimensional subalgebras of~$\mathfrak g$, which then yield systems of ordinary
differential equations. In the following, we classify both subalgebras of the infinite dimensional maximal Lie
invariance algebra $\mathfrak g$ and the subalgebra $\mathfrak g^1$ found in~\cite{sekh12Ay}, as no classification of
one-dimensional subalgebras of $\mathfrak g^1$ was given in~\cite{sekh12Ay}.

The commutation relations between elements of $\mathfrak g$ are
\[
 [\DDD,\GGG]=0,\quad [\LLL(f,g),\DDD]=\LLL(f,g),\quad [\GGG,\LLL(f,g)]=\LLL(f_u,g_u-f),\quad
[\LLL(f_1,g_1),\LLL(f_2,g_2)]=0,
\]
where the pairs $(f,g)$, $(f_1,g_1)$ and $(f_2,g_2)$ are solutions of
system~\eqref{eq:LinearizationModifiedShallowWater}.
The nonidentical basis adjoint actions of the maximal Lie invariance pseudogroup $G$ on the generating vectors fields
$\DDD$, $\GGG$ and $\LLL(f,g)$ then are
\begin{align*}
 &\mathrm{Ad}(e^{\ve \DDD})\LLL(f,g)=e^{\ve}\LLL(f,g),\quad \mathrm{Ad}(e^{\ve \LLL(f,g)})\DDD=\DDD-\ve\LLL(f,g),\\
 &\mathrm{Ad}(e^{\ve \GGG})\LLL(f,g)=\LLL(f',g'),\quad \mathrm{Ad}(e^{\ve \LLL(f,g)})\GGG=\GGG+\ve\LLL(f_u,g_u-f),
\end{align*}
where $f'=f(u-\ve,h)$ and $g'=(g(u-\ve,h)+\ve f(u-\ve,h))$. For each pair of generating vector fields
$(\vv,\mathbf{w}_0)$ of $\mathfrak g$, these adjoint actions can be computed either through the Lie
series~\cite{olve86Ay},
\[
 \mathbf{w}(\ve)=\mathrm{Ad}(e^{\ve \vv})\mathbf{w}_0:=\sum_{n=0}^\infty\frac{\ve^n}{n!}(\mathrm{ad}\
\vv)^n\mathbf{w}_0,
\]
or through a pushforward of the vector fields from $\mathfrak g$ by symmetry transformations from the maximal Lie
invariance pseudogroup $G$, see~\cite{bihl11Dy,card11Ay} for more details. The second method is particularly suitable
for computations involving infinite dimensional Lie algebras.

Now that we have the nonidentical adjoint actions at hand, we can proceed with the classification of the one-dimensional
subalgebras of $\mathfrak g$. To accomplish this classification, we start with the most general element of a
one-dimensional subalgebra from $\mathfrak g$,
\[
 \vv=a\DDD+b\GGG+\LLL(f,g),
\]
where $a,b\in\mathbb{R}$ and $(f,g)$ is an arbitrary but fixed solution of the
system~\eqref{eq:LinearizationModifiedShallowWater}. We subsequently simplify $\vv$ as much as possible by applying the
adjoint actions that we found previously. As the classification is rather short, we will give it explicitly here.

(i) If $a\ne0$, we can scale $a=1$ and use the adjoint action $\mathrm{Ad}(e^{\ve \LLL(\tilde f,\tilde g)})$ to cancel
$\LLL(f,g)$ from~$\vv$. No further simplifications are possible and so the simplified form of $\vv$ is
$\vv=\DDD+b\GGG$.

(ii) For $a=0$ and $b\ne0$, we can set $b=1$ and the adjoint action $\mathrm{Ad}(e^{\ve \LLL(\tilde f,\tilde g)})$ can
be used again to cancel $\LLL(f,g)$, and thus $\vv=\GGG$ in this case.

(iii) The final one-dimensional subalgebra is spanned by $\vv=\LLL(f,g)$. We note that algebras spanned by
$\langle\LLL(f,g)\rangle$ and $\langle\LLL(\tilde f,\tilde g)\rangle$ are equivalent if the pairs of functions $(f,g)$
and $(\tilde f,\tilde g)$ differ by a constant multiplier or a shift of $u$. 

\smallskip

Collecting the results obtained, we have thus proved the following theorem.

\begin{theorem}\label{thm:TheoremOnOneDimensionalAlgebrasModifiedShallowWater}
 An optimal list of inequivalent one-dimensional subalgebras of the maximal Lie invariance algebra $\mathfrak g$ of the
modified shallow-water equations~\eqref{eq:ModifiedShallowWater} consists of the subalgebras
\[
 \langle\DDD+a\GGG\rangle,\quad \langle\GGG\rangle,\quad \langle\LLL(f,g)\rangle,
\]
where $a\in\mathbb{R}$ and $(f,g)$ is an arbitrary but fixed solution of the
system~\eqref{eq:LinearizationModifiedShallowWater}.
\end{theorem}

In principle, Theorem~\ref{thm:TheoremOnOneDimensionalAlgebrasModifiedShallowWater} tells us which
subalgebras should be used for the Lie reductions of system~\eqref{eq:ModifiedShallowWater}. At the same time it is
useful to classify the subalgebra $\mathfrak g^1$ found in~\cite{sekh12Ay}, for which no
optimal system of one-dimensional subalgebras was given. This is a simple task, since the problem of classifying
subalgebras of real four-dimensional Lie algebras is already completely solved~\cite{pate77Ay}. See
also~\cite{muba63Ay,popo03Ay} for classifications of real algebras of dimensions up to six.
For Lie algebras of dimension higher than six exhaustive classification results exist only provided that the Lie algebra
possesses a special structure: see the review in~\cite{popo03Ay} for more details.

In the present case, the algebra $\mathfrak g^1$ is isomorphic to the algebra $A_{4,8}^0$ in the
classification~\cite{popo03Ay}, the nilradical of which is $\langle\p_t,\p_x,\GGG\rangle$, which is isomorphic to the
three-dimensional Heisenberg algebra. The one-dimensional subalgebras of the algebra $A_{4,8}^0$ can be presented as
follows in a way suitable for Lie reduction: 
\begin{equation}\label{eq:OptimalListSubalgebrasSubalgebraModifiedShallowWater}
 \langle\DDD+a\GGG\rangle,\quad \langle\GGG+\delta\p_t\rangle,\quad \langle\p_t+\delta\p_x\rangle,\quad
\langle\p_x\rangle,
\end{equation}
where $a\in\mathbb{R}$ and $\delta\in\{0,1\}$. It is important to note that we cannot set $\delta=0$ in the second
algebra of the above list as the internal equivalence in $\mathfrak g^1$ is weaker than the equivalence in $\mathfrak
g$.

\section{Exact solutions of the modified shallow-water equations}\label{sec:ExactSolutionsModifiedShallowWater}

We now present the associated Lie reductions obtained using the optimal list of one-dimensional
subalgebras of $\mathfrak g$ given in Theorem~\ref{thm:TheoremOnOneDimensionalAlgebrasModifiedShallowWater}. We
solve the arising systems of ordinary differential equations whenever possible.

\emph{(i) Subalgebra} $\langle\DDD+a\GGG\rangle$. The reduction ansatz in this case is $u=\tilde u(p)+a\ln t$, $h=\tilde
h(p)$, where $p=x/t-a\ln t+a$ is the new independent variable. Plugging this ansatz into
system~\eqref{eq:ModifiedShallowWater} reduces the two equations to the system of nonlinear ordinary differential
equations
\[
 a-p\tilde u'+\tilde u\tilde u'+\left(1+\frac{H}{\tilde h}\right)\tilde h'=0,\quad
 -p\tilde h'+\tilde u\tilde h'+\tilde h\tilde u'=0,
\]
where here and in the following a prime denotes the derivative with respect to $p$. As this equation is still too
complicated to be solved in closed form, one could resort to numerical integration for finding a solution of the above
nonlinear system and then extending the numerical solution to a solution of the original modified shallow-water
equations.

\smallskip

\emph{(ii) Subalgebra} $\langle\GGG\rangle$. The ansatz for Lie reduction in this case is $u=\tilde u(p)+x/t$, $h=\tilde
h(p)$, where $p=t$. The system of modified shallow-water equations~\eqref{eq:ModifiedShallowWater} then reduces to
\[
 \tilde u'+\frac{\tilde u}{p}=0,\quad \tilde h'+\frac{\tilde h}{p}=0.
\]
 The solution of this system is $\tilde u=c_1/p$ and $\tilde h=c_2/t$, where $c_1,c_2\in\mathbb{R}$, giving rise to the
Galilean invariant solution
\begin{equation}\label{eq:GalileanInvariantSolutionModifiedShallowWater}
 u=\frac{x+c_1}{t},\quad h=\frac{c_2}{t}
\end{equation}
of the modified shallow-water equations~\eqref{eq:ModifiedShallowWater}.

\smallskip

\emph{(iii) Subalgebra} $\langle\LLL(f,g)\rangle$. For this algebra a suitable reduction ansatz is $u=\tilde u(p)$,
$h=\tilde h(p)$, where $p=fx-gt$. Implicitly differentiating this ansatz with respect to $t$ and $x$ and solving the
resulting algebraic system for the required derivatives $u_t$, $u_x$, $h_t$ and $h_x$ yields
\[
 u_t=-\frac{g\tilde u_p}{D},\quad u_x=\frac{f\tilde u_p}{D},\quad h_t=-\frac{g\tilde h_p}{D},\quad h_x=\frac{f\tilde
h_p}{D},
\]
where the precise expression for $D$ is not required. Plugging this ansatz into the modified shallow-water equations
leads to
\[
 -g\tilde u_p+f\tilde u\tilde u_p+f\left(1+\frac{H}{\tilde h}\right)\tilde h_p=0,\quad -g\tilde h_p+f\tilde u\tilde h_p
+ f\tilde h\tilde u_p=0,
\]
where $f=f(\tilde u,\tilde h)$ and $g=g(\tilde u,\tilde h)$. The reduction with respect to the algebra
$\langle\LLL(f,g)\rangle$ includes several physically relevant solution ansatzes. For example, in the case of $f=1$ and
$g=\const$ (which obviously is a solution of the system~\eqref{eq:LinearizationModifiedShallowWater} as required), the
reduction ansatz coincides with that of a traveling wave solution. Similarly, for $(f,g)=(1,0)$ (or $(f,g)=(0,1)$) we
obtain the stationary (or space independent) reduced system. In the present case, all these reduced
equations admit only the constant solution $u=c_1$ and $h=c_2$ and thus are not of great interest. Less trivial but
implicit solutions can be found by choosing $f$ and $g$ in a way such that they explicitly depend on $u$ and $h$.

\smallskip

We now review succinctly the Lie reductions and results obtained in~\cite{sekh12Ay}. As noted in
the previous section, no optimal list of one-dimensional subalgebras of their algebra $\mathfrak g^1$ was used to
systematically carry out reductions. This gave rise to overly complicated reduction ansatzes
in~\cite{sekh12Ay}. Moreover, these ansatzes could lead to solutions which in fact can be
mapped to each other using a symmetry transformation, i.e.\ that are equivalent.

Three Lie reductions were presented in~\cite{sekh12Ay}. The first reduction (Case Ia) is carried out with respect to the
algebra $\langle a_1\DDD+a_2\GGG+a_3\p_x+a_4\p_t\rangle$, where $a_1\ne0$. In view of our classification results and the
optimal system~\eqref{eq:OptimalListSubalgebrasSubalgebraModifiedShallowWater}, it is obvious that this case is
equivalent to our case (i), i.e.\ the algebra $\langle\DDD+a\GGG\rangle$. Thus, the ansatz found in~\cite{sekh12Ay}
should reduce to the ansatz presented above for case (i) when $a_1=1$ and $a_3=a_4=0$. However, this is not the case. In
fact, it can be directly checked that the ansatz in~\cite{sekh12Ay} for this case is incorrect. The correct (but overly
complicated) ansatz for the algebra $\langle a_1\DDD+a_2\GGG+a_3\p_x+a_4\p_t\rangle$ would be
\[
 p=\frac{a_1^2x+a_1a_3-a_2a_4}{a_1^2(a_1t+a_4)}-\frac{a_2}{a_1^2}\ln(a_1t+a_4),\quad u=\tilde
u(p)-\frac{a_2}{a_1}\ln(a_1t+a_4),\quad h=\tilde h(p).
\]

The second reduction (Case Ib) rests on the algebra $\langle a_2\GGG+a_3\p_x+a_4\p_t\rangle$. It is equivalent to our
case (ii), i.e.\ it is needless to assume $a_3\ne0$, $a_4\ne0$. However, it is crucial to point out that one cannot put
$a_4=0$ using only the weaker equivalence relation imposed by the adjoint action associated with the subgroup $G^1$ of
$G$ on the subalgebra $\mathfrak g^1$. This is why it is instructive to compare the second case in the optimal list of
subalgebras of the maximal Lie invariance algebra~$\mathfrak g$ given in
Theorem~\ref{thm:TheoremOnOneDimensionalAlgebrasModifiedShallowWater} with the second case in the optimal list of
subalgebras of the subalgebra~$\mathfrak g^1$ in~\eqref{eq:OptimalListSubalgebrasSubalgebraModifiedShallowWater} to each
other. On the other hand, even with the adjoint action of $G^1$ on $\mathfrak g^1$, it would be possible to set $a_3=0$.

The last reduction (Case II) employs the algebra $\langle a_2\GGG+a_3\p_x\rangle$. The associated group-invariant
solution is equivalent to our case (ii) i.e.\ it is again needless to assume $a_3\ne0$, and the exact solution found
in~\cite{sekh12Ay} can be obtained from the Galilean invariant
solution~\eqref{eq:GalileanInvariantSolutionModifiedShallowWater} through re-scaling of $t$ and $x$ and shifting of $t$.

To sum up, the three reductions carried out in~\cite{sekh12Ay} are either incorrect (Case Ia) or equivalent to a
solution that could be obtained from a simpler reduction ansatz (Case Ib and Case II). From the more general point of
view, if one would seek only reductions based on the subalgebra $\mathfrak g^1$, the optimal
list~\eqref{eq:OptimalListSubalgebrasSubalgebraModifiedShallowWater} should be used to find the proper reduction
ansatzes.

\begin{remark}
 Although the general strategy for the construction of invariant solutions is to use inequivalent subalgebras to derive
inequivalent solutions, the procedure of generating new solutions from known ones for
system~\eqref{eq:ModifiedShallowWater} could yield interesting results as well since this procedure for
system~\eqref{eq:ModifiedShallowWater} can involve the hodograph transformation. This is why even the use of equivalent
subalgebras independently will make sense for this system if one is able to construct explicit solutions and avoiding
the use of the hodograph transformations in finding these solutions. On the other hand, in~\cite{sekh12Ay} this
construction would be trivial as the transformations considered in~\cite{sekh12Ay} are projectable.
\end{remark}

So far, we have not used the property that~\eqref{eq:ModifiedShallowWater} is linearized
to~\eqref{eq:LinearizationModifiedShallowWater} by the hodograph transformation. We now investigate the possibility of
finding exact solutions of~\eqref{eq:LinearizationModifiedShallowWater} that can be used to obtain exact solutions of
the initial system of modified shallow-water equations~\eqref{eq:ModifiedShallowWater}. To this end, we start by
combining the two equations of the system~\eqref{eq:LinearizationModifiedShallowWater} to a single equation for $f$ by
excluding $g$, which reads
\begin{equation}\label{eq:LinearizationModifiedShallowWaterSingle}
 2f_h+hf_{hh}-\left(1+\frac{H}{h}\right)f_{uu}=0.
\end{equation}
A solution for this equation can be found from the separation ansatz $f=f_1(u)f_2(h)$, which yields the following form
for $f$,
\[
 f=\frac{1}{\sqrt{h}}\left(c_1\sin\left(\sqrt{\frac{c}{H}}u\right)+c_2\cos\left(\sqrt{\frac{c}{H}}
u\right)\right)\left(c_3J_b\left(2\sqrt{c\frac{h}{H}}\right)+c_4Y_b\left(2\sqrt{c\frac{h}{H}}\right)\right),
\]
where $c>0$, $c_1,\dots,c_4\in\mathbb{R}$, $J_b$ and $Y_b$ are the Bessel functions of the first and second kind,
respectively, and $b=\sqrt{1-4c}$. As it is quite intricate to obtain the corresponding solution for $g$ from the
system~\eqref{eq:LinearizationModifiedShallowWater} using the above solution for $f$, we restrict ourselves to finding
particular solutions of system~\eqref{eq:LinearizationModifiedShallowWater} following from simpler forms of $f$.

To give one example, let us set $c=3/16$, $c_2=c_4=0$. The above solution then reduces to
\[
 f=\frac{c_1}{h^{3/4}}\sin\left(\sqrt{\frac{c}{H}}u\right)\sin\left(\sqrt{4c\frac{h}{H}}\right),
\]
and the solution for $g$ from the system~\eqref{eq:LinearizationModifiedShallowWater} becomes
\begin{align*}
 &g=\frac{c_1H}{h^{5/4}}\Bigg(\sqrt{\frac{h}{H}}\sin\left(\sqrt{c\frac{h}{H}}\right)\Bigg(\frac{1}{\sqrt{3}}
\cos\left(\sqrt{4c\frac{h}{H}}u\right)+{}\\
 &\quad+{}\frac{u}{\sqrt{H}}\sin\left(\sqrt{4c\frac{h}{H}}\right)\Bigg)+\frac{h}{H}\cos\left(\sqrt{c\frac{h}{H}}
u\right)\cos\left(\sqrt{4c\frac{h}{H}}\right)\Bigg)+c_5,
\end{align*}
where $c_5\in\mathbb{R}$. This solution cannot be solved for $u$ and $h$ in terms of $f=t$ and $g=x$ globally, but could
be solved (e.g.\ numerically) pointwise and would then yield a solution manifold for the original modified shallow-water
equations. This solution is obviously inequivalent to the group-invariant solutions found above.

Another, somewhat simpler solution of Eq.~\eqref{eq:LinearizationModifiedShallowWaterSingle} is
\[
 f=c_1\frac{u}{h}+c_2\frac{1}{h}+c_3u+c_4
\]
which, when substituted back into~\eqref{eq:LinearizationModifiedShallowWater} yields the following solution for
$g$,
\[
 g=c_1\left(\frac{u^2}{h}+\frac{H}{h}-\ln h\right)+c_2\frac{u}{h}+c_3\left(\frac12u^2-H\ln h-h\right)+c_5
\]
where again $c_1,\dots,c_5\in\mathbb{R}$. In the special case when $c_1=c_3=c_4=c_5=0$, this solution becomes the
Galilean invariant solution~\eqref{eq:GalileanInvariantSolutionModifiedShallowWater} of the modified shallow-water
equations found above. If all the constants $c_1,\dots,c_5$ are non-zero, it is again impossible to solve the above
solutions for $u$ and $h$ globally in terms of $f=t$ and $g=x$.

In the same way one could proceed to find particular solutions of the
system~\eqref{eq:LinearizationModifiedShallowWater} and then extend them to solution of the original modified
shallow-water equations. We will not pursue this idea here further though.

\section{Conclusion}\label{sec:ConclusionModifiedShallowWater}

In the present paper we have reconsidered the problem of studying the Lie symmetries of a system of modified
shallow-water equations~\eqref{eq:ModifiedShallowWater} recently investigated in~\cite{sekh12Ay}. We have derived that
the maximal Lie invariance algebra of this system is in fact infinite dimensional, containing the algebra found
in~\cite{sekh12Ay} as a subalgebra. The infinite dimensional part of the maximal Lie invariance
algebra~\eqref{eq:MaximalLieInvarianceAlgebraModifiedShallowWater} indicates that the
system~\eqref{eq:ModifiedShallowWater} can be linearized using a hodograph transformation, the result of which is the
system~\eqref{eq:LinearizationModifiedShallowWater}. This is not unexpected, as any homogenous first-order system of
partial differential equations with two dependent and two independent variables can be linearized by a hodograph
transformation, provided that the coefficients in the system only depend on the unknown functions. As in~\cite{sekh12Ay}
only a finite dimensional symmetry algebra is found, the indication, provided by the infinite dimensional maximal Lie
invariance algebra, of the existence of a linearizing transformation is missed.

When it comes to determining exact solutions of the system~\eqref{eq:ModifiedShallowWater} we observe some common errors
in finding exact solutions of partial differential equations in~\cite{sekh12Ay}. These errors are
discussed in~\cite{kudr09Ay,popo10By}. The most crucial point missed is that the algebra $\mathfrak g^1$ is not maximal
as a symmetry algebra of the system~\eqref{eq:ModifiedShallowWater}. Moreover, no optimal list of inequivalent
one-dimensional subalgebras of the algebra $\mathfrak g^1$ was constructed. As the algebra $\mathfrak g^1$ is only
four-dimensional this would have been a simple task as for four-dimensional Lie algebras the problem of classifying all
inequivalent subalgebras is already completely solved. If one omits the construction of such an optimal list of
inequivalent subalgebras prior to carrying out the Lie reductions, then one might end up with unnecessarily complicated
reduction ansatzes. This then leads to needlessly complicated exact solutions for the reduced systems. We observed this
with the exact solutions found in~\cite{sekh12Ay}. In another light, carrying out Lie reductions without reference to
the optimal list of inequivalent subalgebras can lead to equivalent exact solutions, i.e.\ solutions that are related
through a symmetry transformation.

Finally, as the linearizing hodograph transformation was missed in~\cite{sekh12Ay} the authors could not construct exact
solutions of the linearized modified shallow-water equations~\eqref{eq:LinearizationModifiedShallowWater}, some of which
can be mapped to exact (and non-Lie) solutions of the modified shallow-water equations~\eqref{eq:ModifiedShallowWater}.
This is another possible source for finding exact solutions of system~\eqref{eq:ModifiedShallowWater}.

\section*{Acknowledgements}

The authors thank Professor Roman O.\ Popovych for useful discussions and helpful remarks on the manuscript. This
research was supported by the Austrian Science Fund (FWF), project J3182--N13 (AB).


{\footnotesize\itemsep=0ex
\begin{thebibliography}{10}
\providecommand{\url}[1]{\texttt{#1}}
\providecommand{\urlprefix}{URL }
\expandafter\ifx\csname urlstyle\endcsname\relax
  \providecommand{\doi}[1]{doi:\discretionary{}{}{}#1}\else
  \providecommand{\doi}{doi:\discretionary{}{}{}\begingroup
  \urlstyle{rm}\Url}\fi
\providecommand{\eprint}[2][]{\url{#2}}

\bibitem{ibra94Ay}
Ames W.F., Anderson R.L., Dorodnitsyn V.A., Ferapontov E.V., Gazizov R.K.,
  Ibragimov N.H. and Svirshchevskii S.R., \emph{{CRC handbook of Lie group
  analysis of differential equations. Vol. 1. Symmetries, exact solutions and
  conservation laws. Edited by N. H. Ibragimov}}, CRC Press, Boca Raton, 1994.

\bibitem{andr98Ay}
Andreev V.K., Kaptsov O.V., Pukhnachov V.V. and Rodionov A.A.,
  \emph{Applications of group-theoretical methods in hydrodynamics}, Kluwer,
  Dordrecht, 1998.

\bibitem{andr88Ay}
Andreev V.K. and Rodionov A.A., {Group analysis of equations of planar flows of
  an ideal fluid in Lagrangian coordinates}, \emph{Dokl. Akad. Nauk SSSR}
  \textbf{298} (1988), 1358--1361.

\bibitem{bihl11Dy}
Bihlo A., Dos Santos Cardoso-Bihlo E.M. and Popovych R.O., Complete group
  classification of a class of nonlinear wave equations, \emph{J. Math. Phys.}
  \textbf{53} (2012), 123515, 32 pp., arXiv:1106.4801.

\bibitem{bihl09Ay}
Bihlo A. and Popovych R.O., Lie symmetries and exact solutions of the
  barotropic vorticity equation, \emph{J.~Math. Phys.} \textbf{50} (2009),
  123102 (12 pages), arXiv:0902.4099.

\bibitem{bihl11Cy}
Bihlo A. and Popovych R.O., Point symmetry group of the barotropic vorticity
  equation, in \emph{{Proceedings of 5th Workshop ``Group Analysis of
  Differential Equations \& Integrable Systems'' (June 6--10, 2010, Protaras,
  Cyprus)}}, 2011 pp. 15--27.

\bibitem{bihl12By}
Bihlo A. and Popovych R.O., Invariant discretization schemes for the
  shallow-water equations, \emph{SIAM J. Sci. Comput. (accepted for
  publication)}  (2012), arXiv:1201.0498, 27 pp.

\bibitem{bihl11Ay}
Bihlo A. and Staufer J., {Minimal atmospheric finite-mode models preserving
  symmetry and generalized Hamiltonian structures}, \emph{Physica D}
  \textbf{240} (2011), 599--606, arXiv:0909.1957.

\bibitem{blum89Ay}
Bluman G. and Kumei S., \emph{{Symmetries and differential equations}},
  Springer, New York, 1989.

\bibitem{blum10Ay}
Bluman G.W., Cheviakov A.F. and Anco S.C., \emph{Application of symmetry
  methods to partial differential equations}, Springer, New York, 2010.

\bibitem{ches09Ay}
Chesnokov A.A., {Symmetries and exact solutions of the rotating shallow-water
  equations}, \emph{Eur. J. Appl. Math.} \textbf{20} (2009), 461--477.

\bibitem{card11Ay}
Dos Santos Cardoso-Bihlo E.M., Bihlo A. and Popovych R.O., {Enhanced
  preliminary group classification of a class of generalized diffusion
  equations}, \emph{Commun. Nonlinear Sci. Numer. Simulat.} \textbf{16} (2011),
  3622--3638, arXiv:1012.0297.

\bibitem{card12Ay}
Dos Santos Cardoso-Bihlo E.M. and Popovych R.O., Complete point symmetry group
  of the vorticity equation on a rotating sphere, \emph{J. Engrg. Math. (in
  print)}  (2012), arXiv:1206.6919, 8 pp.

\bibitem{fush94Ay}
Fushchych W.I. and Popovych R.O., Symmetry reduction and exact solutions of the
  {N}avier--{S}tokes equations, \emph{J. Nonlinear Math. Phys.} \textbf{1}
  (1994), 75--113,156--188, arXiv:math-ph/0207016.

\bibitem{gala06Ay}
Galaktionov V.A. and Svirshchevskii S.R., \emph{Exact solutions and invariant
  subspaces of nonlinear partial differential equations in mechanics and
  physics}, Chapman \& Hall/CRC, Boca Raton, FL, 2006.

\bibitem{hema02Ay}
Hematulin A. and Meleshko S.V., Rotationally invariant and partially invariant
  flows of a viscous incompressible fluid and a viscous gas, \emph{Nonlinear
  Dynam.} \textbf{28} (2002), 105--124.

\bibitem{huar10Ay}
Huard B., {Conditionally invariant solutions of the rotating shallow water wave
  equations}, \emph{J. Phys. A} \textbf{43} (2010), 235205 (20 pages).

\bibitem{hydo00By}
Hydon P.E., How to construct the discrete symmetries of partial differential
  equations, \emph{Eur. J. Appl. Math.} \textbf{11} (2000), 515--527.

\bibitem{hydo00Ay}
Hydon P.E., \emph{Symmetry methods for differential equations}, Cambridge
  University Press, Cambridge, 2000.

\bibitem{ibra95Ay}
Ibragimov N.H., Aksenov A.V., Baikov V.A., Chugunov V.A., Gazizov R.K. and
  Meshkov A.G., \emph{{CRC handbook of Lie group analysis of differential
  equations. Vol. 2. Applications in engineering and physical sciences. Edited
  by N. H. Ibragimov}}, CRC Press, Boca Raton, 1995.

\bibitem{kudr09Ay}
Kudryashov N.A., Seven common errors in finding exact solutions of nonlinear
  differential equations, \emph{Commun. Nonlinear Sci. Numer. Simulat.}
  \textbf{14} (2009), 3507--3529.

\bibitem{levi89By}
Levi D., Nucci M.C., Rogers C. and Winternitz P., {Group theoretical analysis
  of a rotating shallow liquid in a rigid container}, \emph{J. Phys. A}
  \textbf{22} (1989), 4743--4767.

\bibitem{mele04Ay}
Meleshko S.V., A particular class of partially invariant solutions of the
  {N}avier--{S}tokes equations, \emph{Nonlinear Dynam.} \textbf{36} (2004),
  47--68.

\bibitem{mele05Ay}
Meleshko S.V., \emph{Methods for constructing exact solutions of partial
  differential equations}, Mathematical and analytical techniques with
  applications to engineering, Springer, New York, 2005.

\bibitem{muba63Ay}
Mubarakzyanov G.M., {On solvable Lie algebras}, \emph{Izv. Vyssh. Uchebn.
  Zaved. Mat.} \textbf{32} (1963), 114--123, {(in Russian)}.

\bibitem{olve86Ay}
Olver P.J., \emph{{Application of Lie groups to differential equations}},
  Springer, New York, 2000.

\bibitem{ovsi82Ay}
Ovsiannikov L.V., \emph{Group analysis of differential equations}, Acad. Press,
  New York, 1982.

\bibitem{pate77Ay}
Patera J. and Winternitz P., {Subalgebras of real three and four-dimensional
  Lie algebras}, \emph{J. Math. Phys.} \textbf{18} (1977), 1449--1455.

\bibitem{popo99Ay}
Popovych G.V. and Popovych R.O., Flows of incompressible fluids with linear
  vorticity, \emph{J. Appl. Math. Mech.} \textbf{63} (1999), 369--374.

\bibitem{popo00Ay}
Popovych H.V., {L}ie, partially invariant and nonclassical submodels of the
  {E}uler equations, in \emph{Proceedings of Institute of Mathematics of NAS of
  Ukraine}, vol. 43/1, Kyiv, vol. 43/1, 2002 pp. 178--183.

\bibitem{popo95Ay}
Popovych R.O., On {L}ie reduction of the {N}avier--{S}tokes equations, \emph{J.
  Nonlinear Math. Phys.} \textbf{2} (1995), 301--311.

\bibitem{popo03Ay}
Popovych R.O., Boyko V.M., Nesterenko M.O. and Lutfullin M.W., {Realizations of
  real low-dimensional Lie algebras}, \emph{J. Phys. A} \textbf{36} (2003),
  7337--7360, see arXiv:math-ph/0301029v7 for an extended and revised version.

\bibitem{popo10By}
Popovych R.O. and Vaneeva O.O., {More common errors in finding exact solutions
  of nonlinear differential equations: Part I}, \emph{Commun. Nonlinear Sci.
  Numer. Simul.} \textbf{15} (2010), 3887--3899, arXiv:0911.1848v2.

\bibitem{sekh12Ay}
Raja~Sekhar T. and Sharma V.D., Similarity analysis of modified shallow water
  equations and evolution of weak waves, \emph{Commun. Nonlinear Sci. Numer.
  Simulat.} \textbf{17} (2012), 630--636.

\bibitem{vu12Ay}
Vu K.T., Jefferson G.F. and Carminati J., {Finding higher symmetries of
  differential equations using the MAPLE package DESOLVII}, \emph{Comput. Phys.
  Comm.} \textbf{183} (2012), 1044--1054.

\end{thebibliography}

}
\end{document}